\documentstyle[12pt]{article}
\begin{document}
\begin{titlepage}
\begin{center}
\hfill VU-04/94\\
\hfill UUITP $4$/1994\\
\hfill hep-th/9405102\\
\rule[.1in]{13.5cm}{.01in} \\
\vspace{40mm}{\Large\bf Special functions as structure constants \\
for new infinite-dimensional algebras }\\[.4in]
{\large  P.Demkin}\footnote{On leave from Department of Physics,
 Vilnius University, Saul\.{e}tekio al.9, 2054,\\ Vilnius, Lithuania}\\
\bigskip {\it
Institute of Theoretical Physics \\
Uppsala University\\
Box 803, S-75108\\
Uppsala, Sweden\\
paul@rhea.teorfys.uu.se}

\end{center}
\bigskip \bigskip

\begin{abstract}
Novel infinite-dimensional algebras of the Virasoro/Kac-Moody/
\-Floratos-Iliopoulos type are introduced, which involve special
 functions in their structure constants
\end{abstract}

\end{titlepage}
\newpage
\section{Introduction}
Infinite-dimensional algebras of the Virasoro \cite{V}, Kac-Moody
 \cite{V,KM} types have been of increasing \-relevance\- in \-several
 bran\-ches of physics in the last few years and are under intense
 investigation. There are a few reasons to study not only one-loop,
 but also the multi-loop algebras. We have known the two-loop algebra of
 Floratos-Iliopoulos \cite{FI} as an analog of the Virasoro algebra in
 the theory of membranes, correlation with the algebra $W_{\infty }$
 \cite {BK, M, PRS}, connection to the physical states
of the $c=1$ string model \cite {KP, W}. An important role belongs to
 Mojal-Baker-Fairlie algebra \cite {Mo, B, F}, which is connected with the
algebra of area-preserving diffeomorphisms. D.B.Fairlie, P.Fletcher and
 C.K.Zachos some time ago introduced a new type of Lie algebra with
 trigonometric function in structure constants \cite{FFZ}. This algebra
 generalizes or involves as subalgebras the above-mentioned algebras.
 In particular, it can be shown that the classical simple Lie algebras
 $(A_N, B_N, C_N, D_N)$ may be expressed on an "egalitarian" basis
 with trigonometric structure constants \cite{FFZ2}. Here, we introduce
 a new family of infinite-dimensional algebras involving special functions
 in their structure constants, which generalize Fairlie-Fletcher-Zachos
 algebra too.

\section{Virasoro type solutions of Jacobi identity}Consider first the
 general form of algebras on $p$-dimensional integer lattice of indices,
 of the Floratos-Iliopoulos type
\begin{equation}
[L_m, L_n]=f(mn)L_{m+n}\, ,
\end{equation}
where $m=(m_1, m_2, ..., m_p), \, mn=A^{ij}m_i n_j \, , A^{ij}=-A^{ji}$
 - antisymmetric matrix, $f(mn)=-f(-mn)$.

The Jacobi identity dictates the relation for the antisymmetric structure
 constants:
\begin{equation}
f(mn)f(mp+np)+f(np)f(nm+pm)+f(pm)f(pn+mn)=0\, ,
\end{equation}
which is solved by the linear function $f(mn)=rmn+c$. After redefining
 and shifting we have Hoppe algebra \cite{H}. But Jacobi identity admits
other solutions as well $f(mn)$
\begin{equation}
a)  r\sin{(kmn)} \quad b)r\sinh{(kmn)}
\end{equation}
and
\begin{equation}
rcos(kmn)
\end{equation}
at the additional condition $mn + pm + pn ={\pi \over 2} \pm \pi k, k\in Z $
and any function f(mn) at the additional conditions $mn=mp=np$
 and $f(0)=0$; $r,k \in \sc C$ are some arbitrary constants, $r$ to be
 specified by one's convenient normalization of the generators.

\section{Central extension and supersymmetric \, generalization}
The ensuing algebras which also satisfy the Jacobi identity admit
 central extension in the form
\begin{equation}
[L_m, L_n]=f(mn)L_{m+n} + am\delta_{m+n, 0} \, ,
\end{equation}
where $a$ is an arbitrary $p$-vector.

The Fairlie-Fletcher-Zachos algebra \cite{FFZ} is recovered at the
 $A_{ij}=\epsilon_{ij}$ in $p=2$. The Hoppe algebra \cite{H} also
 appears when $f(mn)={1 \over k}mn+c$ in the limit $k\rightarrow 0$.

The supersymmetric extensions of type (1) algebras
\begin{equation}
[L_m, L_n]=f(mn)L_{m+n}  \qquad \{F_m, F_n\}=g(mn)L_{m+n}

\end{equation}
and
\begin{equation}
[L_m, F_n]=f(mn)F_{m+n},
\end{equation}
where $F_n$ are fermionic generators.

For the structure constants $f(mn)$ (3) corresponding antisymmetric
 structure constants $g(mn)$ are
\begin{equation}
a) s\cos{(kmn)} \qquad b)s\cosh{(kmn)}
\end{equation}
and
\begin{equation}
a) s\sin{(kmn)}

\end{equation}
at the additional condition $mn + pm + pn ={\pi \over 2} \pm \pi k,
 k\in Z $ and any function g(mn) at the additional conditions
 $mn=mp=np$ and $g(0)=0$

\section{Lattice average operators and $su(N)$ ideals} Consider the
 features of $2$-dimensional lattice algebras (1). An interesting
 family of algebras with structure constants (3) and (4) and $A_{ij}
\equiv \epsilon_{ij}$ - antisymmetric tensor is the one for which
 $k=2\pi/M$ , for integer $M > 2$. In this family, the 2-dimensional
 lattice separates into fundamental $M^2$ cells. The fundamental
 cell involves only $M^2-1$ points, and there are no more structure
 constants but those occurring in this cell. The sets of "lattice
 average" operators ${\cal L}_m =\sum_{q}{L_{m+Mq}}$, where
 $q=(q_1, ..., q_d)$ belongs to cell, in fact describe $su(M)$ algebras
 for $M$ odd and $su(M/2)$ algebras for $M$ even \cite{FFZ}.

The basis for $su(N)$ algebras with odd $N$ built from two unitary matrices
\begin{eqnarray}
g_o\equiv \left (\matrix{
1 & 0 & 0 & \vdots & 0 \cr
0 & \omega & 0 & \vdots & 0 \cr
0 & 0 & \omega ^2 & \vdots & 0 \cr
\vdots & \vdots & \vdots & \ddots & \vdots \cr
0 & 0 & 0 & \vdots & \omega ^{N-1}
}\right )\quad , \qquad h_0\equiv
\left (\matrix{
0 & 1 & 0 & \vdots & 0 \cr
0 & 0 & 1 & \vdots & 0 \cr
\vdots & \vdots & \vdots & \ddots & \vdots \cr
0 & 0 & 0 & \vdots & 1 \cr
1 & 0 & 0 & \vdots & 0 \cr
}\right ) \\
g_0^N=h_0^N=1\quad , \qquad h_0g_0=\omega g_0h_0\quad,
\end{eqnarray}
where $\omega $ is an $Nth$ root of unity with period no smaller than $N$,
 such as $exp(2\pi i/N)$. The complete set of unitary $N\times N$ matrices
\begin{eqnarray}
J_m=\omega^{m_1m_2/2}g_0^{m_1}h_0^{m_2}, \quad J^\dagger_m=
J_{-m}\\
TrJ_m=0\quad, \qquad m\neq0(modN)
\end{eqnarray}
span the algebra of $su(N)$ and satisfy the algebra
\begin{equation}
[J_m, J_n]=-2i\sin{(2\pi/N)m\times n}J_{m+n},
\end{equation}

For even $N$, the fundamental matrices in (9) are
\begin{eqnarray}
g_e\equiv\sqrt{\omega}g_0 \quad , h_e\equiv

\left (\matrix{
0 & 1 & 0 & \vdots & 0 \cr
0 & 0 & 1 & \vdots & 0 \cr
\vdots & \vdots & \vdots & \ddots & \vdots \cr
0 & 0 & 0 & \vdots & 1 \cr
-1 & 0 & 0 & \vdots & 0 \cr
}\right )\quad , g_e^N=h_e^N=-1 \quad ,
\end{eqnarray}
and the unitary basis is $J_m=\omega^{m_1m_2/2}g_e^{m_1}h_e^{m_2}$
 and $su(N)$ algebra is now
\begin{equation}
[J_m, J_n]=-2i\sin{(\pi/N)m\times n}J_{m+n},
\end{equation}

For nonperiodical structure constants like (3b) we may introduce
 $N\times N$ nonunitary matrices
\begin{equation}
J_m=\omega^{im_1m_2/2}g_0^{im_1}h_0^{im_2},
\end{equation}
which satisfy the algebra
\begin{equation}
[J_m, J_n]=-\sinh{(\pi/N)m\times n}J_{m+n},
\end{equation}
but they do not span the algebra of $su(N)$.

Let $f(n) : \cos{(2\pi/N)m\times n}=-2f(n)f(m)f^{-1}(n+m)$, and $J_n/f(n)$,
 where $J_n$ are generators of algebra (1)(3a):
\begin{equation}
[J_m, J_n]=r\sin{(kmn)}J_{m+n}
\end{equation}
Then
\begin{eqnarray}
J_m=f^{-1}
(m)\omega^{m_1m_2/2}g_0^{m_1}h_0^{m_2}, \quad J^\dagger_m=
J_{-m}\\
TrJ_m=0\quad, \qquad m\neq 0(modN)
\end{eqnarray}
span the algebra of $su(N)$ and satisfy the algebra
\begin{equation}
[J_m, J_n]=-2i\tan{(2\pi/N)m\times n}J_{m+n},
\end{equation}

Correspondingly, the generators

\begin{equation}
J_m=f^{-i}(m)\omega^{im_1m_2/2}g_0^{im_1}h_0^{im_2}
\end{equation}
satisfy the algebra
\begin{equation}
[J_m, J_n]=-\tanh{(\pi/N)m\times n}J_{m+n},
\end{equation}

\section{The algebra of generalized Weyl-Ordered Operators}
There is a close relation between algebras with trigonometric
 structure constants (3) and generalized Weyl-ordered operators.
 Define $T_{j,m}$ as a fully symmetrized, averaged sum of monomials
 of degree $j$ in operator $P$ and $m$ in operator $Q$, i.e. may be
 derived from the generating function
\begin{equation}
\sum^{s}_{j=o}{\left (\matrix{
s\cr
j\cr
}\right )a^jb^{s-j}T_{j,s-j}} = (aP + bQ)^s ,\quad aP\equiv a^iP_i\quad ;
\quad  bQ \equiv b^jQ_j,
\end{equation}
where $P_i , Q_j$ satisfy the canonical commutation relation of the
 Heisenberg algebra
\begin{equation}
P_iQ_j - Q_jP_i = i\lambda \delta _{ij}
\end{equation}

Then the operators
\begin{equation}
E_{a,b}={1 \over 2i\lambda }exp \sqrt{2i} (aP+bQ) \quad ,
\end{equation}
obey to the algebra
\begin{equation}
[E_{a,b}, E_{c,d}]={i \over \lambda }\sinh{{[A,B] \over 2}}E_{a+c,b+d}
\quad ,
\end{equation}
where $A\equiv \sqrt{2i} (aP+bQ) , B \equiv \sqrt{2i} (cP+dQ)$ .
 In the case of quantum correlation of relation (26) with the $(ad)\equiv
 a_id^i , (cb) \equiv c_iB^i$ we have:
\begin{equation}
[E_{a,b}, E_{c,d}]=-{1 \over \lambda }\sin{\lambda[(ad)-(cb)]}
E_{a+c,b+d}\quad ,
\end{equation}

\section{Correlation with the area preserving diffeomorphisms}
In the $\Sigma^2$ case there is a nice isomorphism between
 algebra (28) and Moyal bracket algebra \cite{Mo,GF}:
\begin{equation}
[K_f, K_g]=iK_{\sin{\lambda \left\{ f,g \right\}}},
\end{equation}
where
\begin{equation}
K_f={1 \over 2}f(x-i\lambda\partial_y, y+i\lambda \partial_x )
\end{equation}
We may try to find a similar relation in  the multidimensional case.

In  the two-dimensional case $p=2$, to the compact surface $\Sigma^2$ with
 metric $h_{\alpha \beta } $ and unity area
\begin{equation}
\int d^{2}\xi \sqrt{|deth_{\alpha \beta}(\xi )|} = 1
\end{equation}
we may introduce a complete orthonormal basis $Y_I(\xi )$ to harmonic
 decomposition of the surface coordinates $X^{\mu }$:
\begin{equation}
X^{\mu }=\sum_{I}{x^{\mu  I}Y_I(\xi )} .
\end{equation}
Then in this basis the group of area-preserving diffeomorphism is
 \cite{WMN}
\begin{equation}
[Y_A,Y_B]=f_{ABC}Y^C ,
\end{equation}
where
\begin{equation}
f_{ABC}=\int d^{2}\xi \sqrt{|deth_{\alpha \beta}(\xi)|}Y_A(\xi)[Y_B(\xi),
Y_C(\xi )]
\end{equation}
But the structure constants of this representation  depend on surface
 topology. Therefore, to the multidimensional case let us derive a basis-
independent area-preserving diffeomorphism algebra in terms
 of local differential operators.

Let us consider $p$-dimensional surface $\Sigma^p$ with local
 commuting coordinates $x_i$
and $f_j\in C[\Sigma^p]$ as their differentiable functions.
Then the basis-independent realization for the area-preserving
 diffeomorphism generators is
\begin{eqnarray}
L_{\vec{f}}=\left |\matrix{
f_{1;1} & f_{1;2} & \vdots & f_{1;p-1} & \partial _1 \cr
f_{2;1} & f_{2;2} & \vdots & f_{2;p-1} & \partial _2 \cr
\vdots & \vdots & \ddots & \ldots & \vdots \cr
f_{p;1} & f_{p;2} & \vdots & f_{p;p-1} & \partial _p \cr
}\right | \quad ,
\end{eqnarray}
where $f_{i;j}\equiv \partial_j f_i(\Sigma^p)$ so that the generators
 $L_{\vec{f}}$ transform $dx^i$ to
 $dx^i\rightarrow dx^i+\partial _{(i)}a^idx^i$ (no summing), where
 $a^i$ is cofactor $\partial _i$
in $L_{\vec{f}}$ expression. Infinitesimally, this is a canonical
 transformation which preserves the phase-space area element
 $dx_1dx_2...dx_p$. The explicit form of this transformation at $p=3$ is
\begin{eqnarray}
(x_1, x_2, x_3)\rightarrow \left( x_1+\left |\matrix{
f_{1;2} & f_{2;2}\cr
f_{1;3} & f_{2;3} \cr
}\right | , x_2-\left |\matrix{
f_{1;1} & f_{2;1}\cr
f_{1;3} & f_{2;3} \cr
}\right | , x_3+\left |\matrix{
f_{1;1} & f_{2;1}\cr
f_{1;2} & f_{2;2} \cr
}\right |\right)
\end{eqnarray}

The basis-independent realization for the area-preserving diffeomorphism
 generators $L_{\vec{f}}$ obey to the algebra
\begin{equation}
[L_{\vec{f}}, L_{\vec{g}}]=L_{L_{\vec{f}}\vec{g}} - L_{L_{\vec{g}}\vec{f}}
\end{equation}
At $p=2$ this algebra turns to the well-known algebra \cite{FI}
\begin{equation}
[L_f, L_g]=L_{\{f, g\}}, \qquad \{f, g\} \equiv (\partial f/\partial x_1)
(\partial /\partial x_2) - (\partial f/\partial x_2)(\partial g/\partial x_1)
\end{equation}

Thus, at $p>2$ there is no isomorphism between
 algebra of generalized Weyl-ordered operators and the area-preseving
 diffeomorphysm algebra. This close relation appears only at $p=2$.
 In the general case we have the algebra with two terms, and it has a more
 general form than our solutions of Jacobi identity.

\section{Acknowledgements}
Author wants to express his gratitude to the Swedish Institute for
 Grant 304/01 GH/MLH, which gave him the opportunity to enjoy the
 warm hospitality of Prof.Antti Niemi, Doc.Staffan Yngve and all
 members of the Institute of Theoretical Physics, Uppsala
 University.

\end{document}